\documentclass[12pt]{amsart}
\usepackage{amsmath,amssymb,amscd}
\usepackage{tikz}
\usetikzlibrary{matrix,arrows,decorations.pathmorphing}
\usepackage{fancyhdr}
\renewcommand{\a}{\alpha}

\renewcommand{\l}{\lambda}

\newcommand{\cM}{\mathcal M}

\newcommand{\pa}{\partial}

\newcommand{\g}{{\mathfrak g}}
\newcommand{\h}{{\mathfrak h}}

\newcommand{\Diff}{\mathop{\mathrm{Diff}}\nolimits}

\newcommand{\End}{\mathop{\mathrm{End}}\nolimits}
\newcommand{\Aut}{\mathop{\mathrm{Aut}}\nolimits}
\newcommand{\Hom}{\mathop{\mathrm{Hom}}\nolimits}

\newcommand{\CC}{\mathbb C}
\newcommand{\OO}{\mathcal O}
\newcommand{\RR}{\mathbb R}
\newcommand{\cB}{\mathcal B}

\title{Degenerate integrability of quantum spin Calogero--Moser systems.}
\author{N.Reshetikhin}
\address{N.R.: Department of Mathematics, University of California, Berkeley,
CA 94720, USA \& ITMO University, Kronverkskii ave. 49, Saint Petersburg 197101, Russia
\& KdV Institute for Mathematics, University of Amsterdam,
Science Park 904, 1098 XH Amsterdam, The Netherlands.}
\email{reshetik@math.berkeley.edu}
\begin{document}

\maketitle

\begin{abstract}
The main result of this note is the proof of degenerate quantum integrability of
quantum spin Calogero--Moser systems  and the description of the
spectrum of quantum Hamiltonians in terms of the decomposition of
tensor products of irreducible representations of corresponding Lie algebra.
\end{abstract}

\section*{Introduction}

Recall the definition of quantum spin Calogero-Moser systems. Let $\g$ be a simple
finite dimensional Lie algebra. We fix a choice of the Borel subalgebra in $\g$,
which gives the choice of the Cartan subalgebra $\h$.
Let $V_\mu$ be an irreducible finite dimensional
representation of $\g$ with highest weight $\mu$ and $V_\mu[0]$ be zero weight
subspace.

The Hamiltonian for the spin Calogero-Moser system\footnote{Calogero-Moser model was initially discovered
for $G=SL_n$ \cite{C}\cite{M}\cite{Su} for rational and trigonometric potentials. The generalization to other root systems is due to Olshanetsy and Perelomov \cite{OP}}
corresponding to these data is the following differential operator acting on $V_\mu[0]$-valued
functions on the Cartan subgroup $H\subset G$\footnote{See Appendix \ref{CMA} for examples.}:
\begin{equation}
\label{CMH}
H_{CM}=-\Delta +\sum_{\alpha\in \Delta_+} \frac{X_\a X_{-\a}}{(h_{\a/2}-h_{-\a/2})^2}
\end{equation}
Here $\Delta$ is the Laplacian on $\h^*$ corresponding to the
scalar product given by the Killing form, $X_\a$ are root elements of the Chevalley
basis in $\g$, and $h_\a$ is the function on $H$ corresponding to the
root $\a$, $Ad_h(X_\a)=h_\a X_\a$\footnote{ Note that operators $X_{\pm \alpha}$ act on
$V_\mu$ but their product acts on $V_\mu[0]$.}.

There is an extensive literature on both classical and quantum
spin Calogero--Moser systems. The original Calogero--Moser system
corresponds to $sl_n$ and $\mu=(m,0,\dots, 0)$.  For Calogero-Moser system see for example \cite{ABB}\cite{GH}\cite{LX}\cite{KKS}{\cite{KBBT}
\cite{Ku}\cite{W}
and references therein.

Classical degenerate integrability\footnote{ It is also known as non-commutative integrability
or as superintegrability. Though terms {\it superintegrability} and
{\it non-commutative integrability} are more accepted in the literature,
we will use the term {\it degenerate integrability} to avoid confusion with supergeometry
and quantum integrability} generalizes the notion of Liouville
integrability to the case when the dimension of Liouville tori
is less than half of the dimension of the phase space of the system.
The notion goes back to analysis of the Kepler system, also known as the model of hydrogen atom
in the quantum case.
It can be attributed to Pauli \cite{P}. After a number of new examples
discovered in the 60's , see for example \cite{Winter}, it was formulated in
modern mathematical terms by Nekhoroshev in \cite{N}. Then it was used in a
series of papers by Mischenko, Fomenko and co-authors, for details see \cite{FM}. Some of the work
in this direction is summarized in \cite{SInt}.

Instead of Lagrangian fibration by level sets of Poisson commuting Hamiltonians as
in Liouville integrable systems, a degenerately integrable system on a symplectic manifold
$\cM_{2n}$ is defined by two projections
\begin{equation}\label{cds}
\cM_{2n}\to P_{2n-k} \to B_k
\end{equation}
where $P_{2n-k}$ is a Poisson manifold, and the connected components of
generic fibers of the last projection are symplectic leaves in $P_{2n-k}$.
Functions on $B_k$ form a Poisson commutative subalgebra in the algebra of functions on $\cM_{2n}$.
This subalgebra is Poisson central in the Poisson subalgebra of functions on $P_{2n-k}$.
The subalgebra of functions on $B_k$ (the subalgebra of Hamiltonians) is
the Poisson center of the subalgebra of function on $P_{2n-k}$ (the subalgebra of integrals).

Similar to quantization of Liouville integrable systems, a {\it quantization of a degenerate
integrable system} (\ref{cds}) consists of a sequence of three embedded associative algebras:
\[
C_h(\cM_{2n}) \supset C_h(P_{2n-k}) \supset C_h(B_k)
\]
Here $C_h(\cM_{2n})$ is an associative deformation of the algebra of
functions\footnote{ In the real smooth, or real analytic case this is the space smooth, or real analytic functions
respectively. In the complex algebraic setting this is the space of algebraic functions, etc.} on $\cM_{2n}$.
The subalgebra $ C_h(P_{2n-k})$ is an associative deformation of functions on $P_{2n-k}$, and $C_h(B_k)$ is
a commutative algebra which is a deformation of functions on $B_k$. The algebra $C_h(B_k)$ must be the center of
$C_h(P_{2n-k})$. Also, the algebra $C_h(B_k)$ must be the centralizer of $ C_h(P_{2n-k})$  in $C_h(\cM_{2n})$.
The algebra $C_h(P_{2n-k})$ is called the algebra
of quantum integrals, the algebra $C_h(B_k)$ is the algebra of quantum Hamiltonians.

Classical degenerate integrability of spin Calogero--Moser models was proved in \cite{R} where the duality between the spin Calogero-Moser model and the spin Ruijsenaars model was also noted. For latest developments on this duality see \cite{AF}. The main result of this paper is the proof of quantum
degenerate integrability of the quantum spin Calogero--Moser system.  The description
of eigenfunctions and of the spectrum of quantum Hamiltonians can be found in \cite{Et}\cite{FP} respectively.
We also present it here for completeness.

In Section \ref{qdCM} we describe the sequence of algebras responsible for degenerate integrability of the quantum
spin Calogero--Moser systems. Eigenfunctions of quantum spin Calogero--Moser Hamiltonians and described in Section \ref{CM-spec}, where we also describe
the spectrum of quantum Hamiltonians in the compact case. Conclusion contains a few remarks
on semiclassical asymptotic of degenerate integrable systems, a comment on the relation to
earlier works \cite{W} and \cite{Ku}.

\section{Quantum degenerate integrability of spin Calogero-Moser systems.}\label{qdCM}

In this paper we will consider only quantum systems that are quantizations
of classical systems. The core structure of a degenerate
quantum integrable system consists of an algebra $C$ with trivial center
and two subalgebras:
\[
C\supset A \supset B
\]
where $B=Z(A)$ (the center of $A$) and $B=Z(A,C)$ (the centralizer of $A$ in $C$.

\subsection{Algebra of quantum integrals}
Let $\Diff(G)$ be the algebra of differential operators on $G$.
We have a natural embedding of the left- and the right-invariant vector
fields by Lie derivatives. Polynomials in these Lie derivatives
can be identified with two copies of the universal enveloping
algebras $U(\g_L)$ and $U(\g_R)$, where $\g_L$ and $\g_L$ are
two copies of $\g$ realized by right and left Lie derivatives
respectively. It is clear that $Ad_G$-invariant polynomials in left
and right Lie derivatives coincide. Therefore, we have the natural embedding:
\[
\Diff(G)\supset U(\g)\otimes_{Z(U(\g))} U(\g)
\]
Here the first factor in the tensor product corresponds to left-invariant
vector fields and the second factor, to right-invariant vector fields.
The tensor product is taken over the ring of $Ad_G$-invariant polynomials
in Lie derivatives. The space of such polynomials can be naturally identified with the
center of $U(\g)$.

Let $\Diff_G(G)$ be the algebra of differential operators invariant with
respect to the adjoint $G$-action.  Denote by $(U(\g) \otimes_{Z(U(\g))}
U(\g))_G$ the subalgebra of $U(\g) \otimes_{Z(U(\g))} U(\g)$ of
$G$-invariant elements, where $G$ acts diagonally on the tensor product.
We have the following commutative diagram:
\[
\begin{tikzpicture}[scale=1.5]
\node (A) at (0,1) {$\Diff(G)$};
\node (B) at (3,1) {$U(\g)\otimes_{Z(U(\g))}U(\g)$};
\node (F) at (6,1) {$U(\g)$};
\node (C) at (0,0) {$\Diff_G(G)$};
\node (D) at (3,0) {$(U(\g)\otimes_{Z(U(\g))}U(\g))_G$};
\node (E) at (6,0) {$Z(U(\g))$};
\path[left hook->,font=\scriptsize,>=angle 90]
(B) edge (A)
([yshift= 1.5pt]F.west) edge node[above] {$L$} ([yshift= 1.5pt]B.east)
([yshift=-1.5pt]F.west) edge node[below] {$R$} ([yshift=-1.5pt]B.east)
(E) edge node[above] {$i$} (D)
(D) edge (C)
(C) edge (A)
(D) edge (B)
(E) edge (F);
\end{tikzpicture}
\]
Here $L(a)=a\otimes 1$, $R(a)=1\otimes a$  and
the map $i$ comes from the natural isomorphism $Z(U(\g)) \simeq (Z(U(\g)
\otimes_{Z(U(\g))} U(\g)))_G$. It acts as $z\mapsto z\otimes_{Z(U(\g))} 1=1 \otimes_{Z(U(\g))} z$.

This diagram is the quantum version of the corresponding diagram
for Poisson projections which prove the degenerate integrability
of spin Calogero--Moser systems in the classical case.

\begin{equation}\label{D}
\begin{tikzpicture}[scale=1.5]
\node (A) at (0,1) {$T^*G$};
\node (B) at (3,1) {$\g^*\times_{\h^*/W} \g^*$};
\node (F) at (6,1) {$\g^*$};
\node (C) at (0,0) {$T^*G/Ad_G$};
\node (D) at (3,0) {$(\g^*\times_{\h^*/W} \g^*)/G$};
\node (E) at (6,0) {$\h^*/W$};
\path[->,font=\scriptsize,>=angle 90]
(A) edge (B)
([yshift= 1.5pt]B.east) edge node[above] {$L$} ([yshift= 1.5pt]F.west)
([yshift=-1.5pt]B.east) edge node[below] {$R$} ([yshift=-1.5pt]F.west)
(D) edge node[above] {$p$} (E)
(C) edge (D)
(A) edge (C)
(B) edge (D)
(F) edge (E);
\end{tikzpicture}
\end{equation}

Here $\g^*\times_{\h^*/W} \g^*$ is the fibred product of two copies of $\g^*$
over $\h^*$. The maps in the upper row of the diagram act as $(x,g)\mapsto (x, -Ad^*_g(x))$, $L(x,y)=x$, and $R(x,y)=y$. Here and below we assume that the co-adjoint bundle $T^*G$ is trivialized
by left translations $T^*G\simeq \g^*\times G$. The lower horizontal sequence of Poisson maps
lies at the heart of degenerate integrability
of the spin Calogero--Moser systems \cite{R}.

Recall that classical spin Calogero-Moser systems are parameterized by co-adjoint orbits $\OO \subset \g^*$.
If $\OO$ is passing through $t\in \h^*$, then it is also passing through each $w(t)$, where $w\in W$
is an element of the Weyl group. We will denote such orbit passing through $t$ by $\OO_{[t]}$ where $[t]\in \h^*/W$
is the orbit of $t$ with respect to the Weyl group action. When $\g$ is a real compact form of a simple Lie algebra,
we can identify $\h^*/W$ with $h^*{\geq 0}=\{\sum_{i=1}^r x_i\omega_i| x_i\in \RR_{\geq 0}\}$, where $\omega_i$ are the fundamental weights of $\g$, and $r$ is the rank of $\g$.

For a generic co-adjoint orbit $\OO_{[t]}$  the phase space of the corresponding spin Calogero-Moser system is
the symplectic leaf $S_{[t]}=\mu^{-1}(\OO_{[t]})/G$, where $\mu: T^*G\to \h^*$ is the moment map for
the adjoint action of $G$:
\[
\mu(x,g)=x-Ad^*_g(x)\in \g^*
\]
Here $x\in \g^*, g\in G$.

The sequence of projections from the diagram above produces the sequence of
projections
\begin{equation}\label{CM-dint}
S_{[t]}\to \sqcup_{[s]\in \h^*/W} \cM_{[s], -[s]|[t]} \to \cB_{[t]}\subset \h^*/W
\end{equation}
Here the moduli space $\cM_{[s_1], [s_2]|[t]}$ is defined as
\[
\cM_{[s_1], [s_2]|[t]}=\{(x_1,x_2)\in \OO_{[s_1]}\times \OO_{[s_2]}| x_1+x_2\in \OO_{[t]}\}
\]
and $\cB_{[t]}=\{[s]\in \h^*/W| \cM_{[s], -[s]|[t]}\neq \emptyset \}$. Note that
$\cB_{[t]}$ is unbounded but if $t\neq 0$ it does not contain the vicinity of zero.
The series of projections (\ref{CM-dint}) describes the degenerate integrability of classical
spin Calogero-Moser model. The Hamiltonian of the classical spin Calogero-Moser system
is the pull-back of the quadratic Casimir function on $\h^*/W$ to $S_{[t]}$.
Taking into account the isomorphism $S^{reg}_{[t]}\simeq (T^*\h\times \OO_{[t]}//H)/W$, where
$\OO_{[t]}//H$ is the Hamiltonian reduction of $\OO_{[t]}$ with respect to the coadjoint action of
$H$, the Hamiltonian of the classical spin Calogero-Moser system can be written as
\[
H_{CM}=<p,p>+ \sum_{\alpha\in \Delta_+} \frac{x_\a x_{-\a}}{(h_{\a/2}-h_{-\a/2})^2}
\]
where $p, h_\alpha$ are coordinate functions on $T^*\h$ and $x_\a x_{-\a}$ is a function on $\OO_{[t]}//H$,
see \cite{R} for details.

The sequence of embeddings of algebras
\[
\begin{tikzpicture}[scale=1.5]
\node (C) at (0,0) {$\Diff_G(G)$};
\node (D) at (3,0) {$(U(\g)\otimes_{Z(U(\g))}U(\g))_G$};
\node (E) at (6,0) {$Z(U(\g))$};
\path[left hook->,font=\scriptsize,>=angle 90]
(D) edge (C)
(E) edge node[above]{$i$} (D);
\end{tikzpicture}
\]
quantizes the sequence of Poisson projections from the low row of the commutative diagram (\ref{D}).

We will use this sequence to demonstrate degenerate quantum integrability of
quantum spin Calogero--Moser systems. The subalgebra $(U(\g)\otimes_{Z(U(\g))}U(\g))_G\subset \Diff_G(G)$
gives the algebra of quantum integrals, and the subalgebra $Z(U(\g))\subset \Diff_G(G)$ gives
the subalgebra of quantum integrals. The resulting system of embedded algebras will quantize Poisson projections
(\ref{CM-dint}).

\subsection{Quantum Spin Calogero--Moser Hamiltonians}

\subsubsection{} Recall that in order to define quantum spin Calogero--Moser system
we need to fix an irreducible finite dimensional representation of $\g$.
Let $\mu$ be the highest weight of such representation and $V_{\mu}[0]$ be its
zero weight subspace.

Denote by $W_\mu(H)$ the space of $V_{\mu}[0]$-valued smooth functions on the Cartan subgroup $H\subset G$ which are
equivariant with respect to the action of the Weyl group of $G$:
\[
f(wh)=wf(h)
\]
Here the action of $W$ on $V_\mu[0]$ is induced by the action of $N(H)$, the normalizer of $H\subset G$.

We have a natural embedding of $W_\mu(H)$ into the space $C_\mu(G)$ of $V_\mu$-valued smooth functions on $G$.
The image of this imbedding we denote by $W_\mu(G)$:
\[
W_\mu(G)=\{f\in C_\mu(G)|f(g^{-1}hg)=\pi_\mu(g)f(h), \ g\in G, \ h\in H, f(h)\in V_\mu[0]\subset V_\mu\}
\]

For functions from $W_\mu(G)$ we have:
\[
f(g^{-1}g'g)=\pi_\mu(g)f(g')
\]

\subsubsection{} Let $x \in \g$ and $x^L$ and $x^L$ be corresponding left and right
invariant vector fields on $G$ which we identify with the corresponding Lie derivatives.
If $f\in W_\mu(G)$
\begin{equation}\label{eq1}
((x^L + x^R)f)(g) = \pi^{\mu}(x)f(g).
\end{equation}

Let $\{X_{\a}, H_i\}$, where $\a$ are roots, and $i=1,\dots, r=rank(G)$, be a Chevalley basis in $\g$. For the right and left action of these basis elements on functions from $W_\mu(G)$, followed by the restriction to $H\subset G$, we have:
\begin{eqnarray*}
(X_{\a}^Rf)(h) &= &\frac {d}{dt} f(he^{-tX_{\a}})|_{t=0} = \frac {d}{dt}
f(e^{-th_{\a}X_{\a}}h)|_{t=0} = -h_{\a}(X_{\a}^Lf)(h) \\
(H_i^Rf)(h) &= &-(H_i^Lf)(h) = h_i \frac {\pa}{\pa h_i} f(h).
\end{eqnarray*}
Here $h_\a$ the value of $h$ on the root $\a$:
\[
Ad_h(X_\a)=h_\a X_\a
\]
In combination with \eqref{eq1} we have:
\[
((X_{\a}^L - h_{\a}X_{\a}^L)f)(h) = \pi^{\mu}(X_{\a})f(h).
\]
Thus, we have the following explicit action of left and right
invariant vector fields on $G$ on the space $W_\mu(H)$:
\begin{equation}\label{act-H}
(X_{\a}^Lf)(h) = (1-h_{\a})^{-1}\pi^{\mu}(X_{\a})f(h)
\end{equation}
\[
(X_{\a}^Rf)(h) = -h_{\a}(1-h_{\a})^{-1} \pi^{\mu}(X_{\a})f(h)
\]
\[
(H_i^Rf)(h) = -(H_i^Lf)(h) = h_i \frac {\pa}{\pa h_i} f(h).
\]

The algebra $\Diff_G(G)$ of $G$-invariant differential operators on $G$
acts naturally on the space $W_\mu(G)$. Because the algebra $(U(\g) \otimes_{Z(U(\g))} U(\g))_G$
can be identified with a subalgebra of $\Diff(G)$, it acts naturally on the space
$W_{\mu}(G)$.   Let $I_{\mu}$ be the ideal in $(U(\g)) \otimes_{Z(U(\g))} U(\g))_G$ defined by the
highest weight $\mu$ considered as an irreducible $Z(U(\g))$-character $\mu$.
We have the homomorphism of algebras
\[
\phi: (U(\g) \otimes_{Z(U(\g))} U(\g))_G/I_{\mu} \rightarrow \Diff(H)
\otimes \End(V_{\mu}[0])
\]
The formulae (\ref{act-H}) describe this homomorphism explicitly.

Denote the image of this homomorphism $A_{\mu}$.
Images of elements $i(Z(U(\g))$ in $\Diff(H)
\otimes \End(V_{\mu}[0])$ form commutative subalgebra $B_{\mu} \subset
A_{\mu}$.  This commutative subalgebra is generated by images of
Casimir elements (whose degrees are exponents of $\g$). These elements remain
independent in $B_\mu$ for generic $\mu$\footnote{In the semiclassical limit, when $h\to 0$, we have $Spec(B_{t/h})\to \cB_{[t]}$, in the
appropriate sense.}.

The Hamiltonian of quantum spin Calogero--Mose systems is the image of the
quadratic Casimir in $Z(U(\g) \otimes_{Z(U(\g))} U(\g))_G \simeq
Z(U(\g))$:
\[
{\hat H}_{CM}  = \frac {1}{2}\phi \circ i(c_2)
\]
Here $\phi$ is defined above and $i$ is defined in the last diagram of section 1.1.

Quantum degenerate integrability of the spin Calogero-Moser system is given by the sequence of subalgebras
\[
\begin{tikzpicture}
\node (C) at (0,0) {$\Diff(H) \otimes \End(V_{\mu}[0])$};
\node (D) at (3,0) {$A_\mu$};
\node (E) at (6,0) {$B_\mu$};
\path[left hook->,font=\scriptsize,>=angle 90]
(D) edge (C)
(E) edge (D);
\end{tikzpicture}
\]
These subalgebras quantize the system of Poisson projections
(\ref{CM-dint}) ensuring the degenerate integrability of the classical spin Calogero-Moser system.

The algebra $B_\mu$  can be described as follows. Consider canonical elements (mixed Casimirs):
\[
X^L=\sum_iH^i\otimes H^L_i+\sum_\alpha e_\alpha\otimes X^L_\alpha=
-\sum_iH^i\otimes h_i\frac{\pa}{\pa h_i}+\sum_\alpha \frac{X_\alpha\otimes \pi^\mu(X_\alpha)}{1-h_\alpha}
\]
\[
X^R=\sum_iH^i\otimes H^R_i+\sum_\alpha e_\alpha\otimes X^R_\alpha=
\sum_iH^i\otimes h_i\frac{\pa}{\pa h_i}-\sum_\alpha \frac{X_\alpha\otimes \pi^\mu(X_\alpha)h_\alpha}{1-h_\alpha}
\]
The algebra $B\mu$ is generated by elements
\[
H^V_n=(tr_V\otimes id)(X^L)^n)=(tr_V\otimes id)(X^R)^n)
\]
Here we assume that elements are acting on $W_\mu(H)$.

The elements
\[
(tr\otimes id)((X^L)^{n_1}(X^R)^{n_2}(X^L)^{m_1}\dots)
\]
acting on $W_\mu(H)$ are typical elements from $A_\mu$. The exact description
of elements of $A_\mu$ is part of the geometric invariant theory and we will not discuss it here.

\section{Eigenfunctions of quantum Calogero--Moser systems}\label{CM-spec}

In this case the Hamiltonian (\ref{CMH}) can be written in terms of $q$-coordinates, $h=\exp(iq)$, as follows:
\[
H=-\Delta +\frac{1}{4}\sum_{\alpha >0} \frac{X_\alpha X_{-\alpha}}{\sin^2(\frac{q_\alpha}{2})}
\]
where $0<q_{\alpha}<\pi$ are coordinates on $\mathfrak{h}^* \mod \mathbb{Z}^r$, $r=rank(\g)$.
When $G$ is a compact real form of a simple Lie group, $q_\a$ are real. For split real form they are imaginary.
This construction of eigenvectors and of quantum Hamiltonians follows \cite{Et}.

\subsection{Construction of eigenfunctions}
Here we follow \cite{Et}. Let $(L_{\l},\pi^{\l})$  be a representation
of the Lie group $G$ with an irreducible central character $\l$. Here $\pi^\lambda: G \rightarrow \Aut(L_{\l}))$
is the corresponding group homomorphism.  It induces naturally a representation of
the universal enveloping algebra $U(\g)$. Central elements of $U(\g)$ acts as $\pi^\lambda(z)=z(\lambda)I_{V_\lambda}$
where $z(\lambda)$ is the value of $\lambda$ on $z$. Let
${\hat z} \in \Diff_G(G)$ be the
image of $z\in Z(U(\g))$ in the algebra of differential operators on $G$.  Then
\[
({\hat z}\pi^{\l})(g) = z(\l)\pi^{\l}(g)
\]
Thus, $\pi^{\l}$ is a joint eigenfunction of the subalgebra $Z(U(\g)\subset \Diff_G(G)$.

Let $K^\lambda_\mu(a): L_\lambda\to L_\lambda\otimes V_\mu$ be
$G$-linear map. Here we assume that indices $a$ enumerate all such maps, i.e. $a$ enumerates a basis
in $Hom_G(L_\lambda\to L_\lambda\otimes V_\mu)$. Define the function
\begin{equation}\label{eigen-CM}
f_{\l,\mu}(g,a) = tr_{L_\lambda}(K_{\mu}^{\l}(a) \circ \pi_{\l}(g))
\end{equation}
when the trace converges. This function is an element of $W_\mu(G)$ . It is a linear combination of matrix elements
of $\pi^\lambda(g)$ and therefore it satisfies the differential
equation
\[
\hat{z} f_{\l,\mu}(g,a) = z(\l)f_{\l,\mu}(g,a)
\]
for each $z \in Z(U(\g))$. Restricting this to $H\subset G$ we have
\[
\phi\circ i(z) f_{\l,\mu}(h,a) = z(\l)f_{\l,\mu}(h,a)
\]
Here $f_{\l,\mu}(h,a)\in V_\mu[0]$. When $z=c_2$ the operator on the left is the
quantum spin Calogero-Moser Hamiltonian.

Thus we constructed joint eigenfunctions
of $B_\mu$, and in particular, eigenfunctions for the quantum spin Calogero--Moser Hamiltonians.
A generalization of this construction to the case when
conjugation action of $G$ is replaced by a twisted conjugation action was found in \cite{FP}.

\subsection{The spectrum of quantum Hamiltonians for the compact real form of a simple Lie algebras}
In order to construct the spectrum of quantum Hamiltonians we need a Hilbert space
structure on $W_\mu(G)$.
Now let us consider spin  Calogero--Moser systems corresponding to
compact real forms of simple Lie groups. We assume that representations $V_\mu$ and $L_\lambda$
from previous constructions are irreducible finite dimensional representations.

Define the Hilbert space structure on the space $W_\mu(G)$ (on its $L_1$-completion)
by the Hermitian scalar product
\[
(f_1,f_2)=\int_G <\overline{f_1(g)}, f_2(g)> dg
\]
where $<x,y>$ is the Hermitian invariant scalar product on $V_\mu$ and
$dg$ is the Haar measure on $G$.

In what follow both $L_\l$ and $V_\mu$ are finite dimensional irreducible
representations of $G$. We will use two different letters for them to
emphasize their different roles in the construction.

Choose a basis in the space $Hom(L_\lambda^*\otimes L_\lambda\to V_mu)$ enumerated by
indices $a, b, ...$. Choose a basis in $L_\l$ enumerated by $\alpha, \beta,...$.
Let $K^\l_\mu(a)$ be a $G$-linear map corresponding to the index $a$.
Its Hermitian conjugate $(K\l_\mu(a))^*$ is a $G$-linear map $V_\mu\to L_\lambda^*\otimes L_\lambda$  Assume that this basis is orthonormal:

\begin{equation}\label{ort}
K^\l_\mu(a)(K^\l_\mu(a'))^*=\delta_{a,a'}id_{V_\mu}, \ \ \sum_{a,\mu}(K^\l_\mu(a))^*K^\l_\mu(a)=id_{L_\l^*\otimes L_\l}
\end{equation}

The first identity here is orthonormality and the second, is completeness.
The eigenfunctions of Calogero--Moser Hamiltonians constructed in the
previous section can be written as
\[
f_{\l,\mu}(g,a)=<K^\l_\mu(a),\pi^\l(g)>=\sum_{\alpha, \beta}K^\l_\mu(a)_{\alpha,\beta}\pi^\l(g)_{\alpha, \beta}\in V_\mu
\]
Let us check the orthogonality and completeness of eigenfunctions of Calogero--Moser Hamiltonians
constructed in the previous section. We have
\[
\int_G  \sum_{\alpha, \beta,\gamma,\delta}\overline{K^\l_\mu(a)_{\alpha,\beta,i}} K^{\l'}_\mu(a')_{\gamma,\delta,j}
\overline{\pi^\l(g)_{\alpha,\beta}}\pi^\l(g)_{\gamma,\delta} dg=\delta_{\l,\l'}\delta_{a,a'}\delta_{i,j}
\]
Here we used the Schur's lemma and (\ref{ort}), indices $i,j,..$ enumerate a basis in $V_\mu$. This proves the orthogonality. We also have
\[
\sum_\l\sum_a \overline{f_{\l,\mu}(g,a)_i} f_{\l,\mu}(g',a)_j=\delta_{g,g'}\delta_{i,j}
\]
which proves completeness.

Here we used the completeness of the basis of functions $\pi^\l(g)_{\alpha,\beta}$ in $L_2(G)$:
\[
\sum_\l \overline{\pi^\l(g)_{\alpha,\beta}}\pi^\l(g')_{\gamma,\delta}=\delta_{g,g'}\delta_{\alpha,\delta}
\delta_{\beta,\gamma}
\]
and (\ref{ort}).

Thus, in the compact case the system of eigenvectors $f_{\l,\mu}(g,a)$ is orthonormal and
complete. The multiplicity of $\l$'s joint eigenvalue is equal to
$\dim(\Hom(L_{\l}^* \otimes L_{\l} \rightarrow V_{\mu}))$.

Note that in terms of $q$-coordinates the orthogonality of eigenfunctions can be written as
\[
\int_{t^*} \overline{f_{\l,\mu}(e^{iq},a)_i}f_{\l',\mu}(e{iq},b)_j
\prod_{\alpha>0}\left| e^{\frac{iq_\alpha}{2}}-e^{-\frac{iq_\alpha}{2}}\right|^2 dq=\delta_{a,b}\delta_{\l,\l'}\delta_{ij}
\]
Here $dq$ is the Euclidean measure on $t^*=\mathfrak{h}^*\mod \mathbb{Z}^r$ corresponding to
the Killing form, and $i,j$ enumerate a basis in $V_\mu[0]$.
This is an immediate consequence of the "radial" structure
of the Haar measure on simple compact groups.

Now algebras $A_\mu$ and $B_\mu$ can be described in terms of the spectral decomposition:
\[
B_\mu\simeq \oplus_{\lambda\in C_\mu} E_\lambda
\]
where $E_\lambda$ is a complex one-dimensional, and $C_\mu$ is the set of all integral
dominant weights $\lambda$ such that $dim(Hom_G(V_\lambda, V_\lambda\otimes V_\mu))\neq 0$.
The algebra $A_\mu$ has the following decomposition:
\[
A_\mu\simeq \oplus_{\lambda\in C_\mu} End(Hom_G(V_\lambda, V_\lambda\otimes V_\mu))
\]
The decomposition also describes the action of $A_\mu$ on eigenfunctions of quantum Hamiltonians.

\section{Conclusion}

First, let us say few words about semiclassical limit when components of all weights $\l,\mu,...$
go to infinity so that components of weights remain projectively finite, i.e. they
are all proportional to $N\to \infty$ with finite coefficients.  The quantity $1/N$ plays the
role of the Planck constant. In this limit the quantum Calogero--Moser system becomes a classical one.
Note that the multiplicity of eigenvalues grows as a power of $N$. In the appropriate
sense the endomorphism algebra of the multiplicity space becomes the algebra of functions
on the corresponding symplectic leaf of $P_{2n-k}$. In particular, in the compact case
\[
\dim((\Hom(V_{\l}^* \otimes V_{\l} \rightarrow V_{\mu}))= h^{-n/2}vol(\cM)(1+O(h))
\]
when $\mu=t/h, \l=s/h$, $h\to 0$, $s,t\in \h^*_{\geq 0}$, and we identify $\h^*/W$ with $\h^*_{\geq 0}$
by choosing corresponding representatives in $W$-orbits. Here $n$ is the dimension and $vol(\cM)$ is the
symplectic volume of the moduli space
$\cM_{s, -w_0(s)|t}$ with its natural symplectic structure.

This is a common feature that distinguishes Liouville integrable
systems from degenerately integrable systems. In the first case, generic semiclassical
joint eigenstates of Hamiltonians are non-degenerate, while in the second case, they
are either infinitely degenerate (in the non-compact case) or become infinitely degenerate
as $h\to 0$. The algebras of endomorphisms of multiplicity spaces can be regarded as a
quantization of algebras of functions on symplectic leaves of $P_{2n-k}$.

The proof of degenerate integrability of q-spin Calogero--Moser models and corresponding Ruijsenaars type
models is completely parallel. The difference is that is the q-case one has to use quantized universal enveloping algebras. This will be done in a separate publication.

Degenerate classical and quantum integrability of spin Calogero-Moser systems is part of
a more general construction involving affine loop groups. For the elliptic case, see \cite{KBBT}.

We discussed in details the structure of the spectrum in the compact case.
Similarly, one can study non-compact real forms of simple complex algebraic
Lie groups. Eigenfunctions are still given by  (\ref{eigen-CM}) but the
orthogonality and completeness property will be different and will depend on the structure of
corresponding unitary irreducible representations.

Here we made an assumption that $\mu$ is generic. If it is not, i.e. if its stabilizer
is greater then $H$, the dimension of $\cB_{[t]}$ in the classical case drops by the rank of the
stabilizer and the system becomes even more degenerately integrable. For $SL_n$ the "extreme" case is when
the stabilizer is $SL_{n-1}$. In this case our results give the Liouville integrability of the usual Calogero-Moser
system \cite{KKS} and of its quantum counterpart.

The results of this paper can be generalized to spin q-Calogero-Moser systems and
q-Ruijsenaars systems.  For degenerate classical integrability of these systems see \cite{R1}.
The spectrum of corresponding quantum Hamiltonians is constructed in \cite{EV}.

Finally, a few words about the relation to works \cite{W} and \cite{Ku}.
In \cite{W} a maximal degenerate integrability of classical (non-spin) Calogero-Moser
system was established. Essentially, the result of \cite{W} is that any trajectory
of classical Calogero-Moser system is periodic. This was extended to quantum case
in \cite{Ku}. We conjecture that the spin Calogero-Moser systems have the same property.

\section{Acknowledgments} The author is grateful to  Chebyshev
Lab of St.Ptersburg University and St. Petersburg Mathematics Institute (PDMI)
for the hospitality, to the Russian Science Foundation
(project no. 14-11-00598) for the support during the stay in St.Petersburg,
when this paper was completed. This research was also supported by the NSF grant DMS-0901431 and by the Chern-Simons endowment. He thanks L. Feher for discussions and important
remarks and a referee for useful suggestions.

\appendix

\section{More on quantum spin Calogero-Moser systems}\label{CMA}

\subsection{}For $\g=sl_N$, when $\mu$ is a highest weight corresponding to a
partition of $N$, the space $V_\mu[0]$ can be realized as a subspace of the $N$-th tensor power of the $n$-dimensional representation. Here we will remind that construction which is based on Howe and Schur-Weyl dualities (it is well known and details can be found in many textbooks).

The Howe duality states Lie groups $SL_n$ and $SL_N$, acting on the space of polynomials on matrices $n\times N$ in a natural way, centralize each other. Moreover, the subspace of homogeneous polynomials of degree $m$ decomposes
as $SL_n\times SL_N$ module as follows:
\[
Pol(M_{n, N})_m\simeq \oplus_{\mu \in P(m,n)} V_\mu^{SL_n}\times V_\mu^{SL_N}
\]
Here we assume that $n\leq N$, $P(m,n)$ is a set of partitions of $m$ with at most $n$ nonempty rows,
and $V_\mu^{SL_n}$ is an irreducible finite dimensional $SL_n$ module corresponding to the partition $\mu$.

When $m=N$ the subspace
\[
\oplus_{\mu\in P(N,n)} V_\mu^{SL_n}\times V_\mu^{SL_N}[0]
\]
corresponds to the subspace of $Pol(M_{n, N})_N$ spanned by monomials of the form $M_{\alpha i}$
where each $i=1,\dots, N$ appear exactly once. This subspace can be identified with $(\CC^n)^{\otimes N}$
as follows. We identify monomials with the tensor product basis as
\[
M_{\alpha_1, 1}\dots M_{\alpha_N, N} \mapsto e_{\alpha_1}\otimes \dots \otimes e_{\alpha_N}
\]
The group $SL_N$ does not act anymore on $V_\mu^{SL_N}[0]$, its Cartan subgroup $H$ acts trivially on it, and the
normalizer of the Cartan $N(H)\subset SL_N$ acts by permutations of indices $\alpha_i$. In other words,
it induces the action of
the symmetric group $S_N$ on $V_\mu[0]$, which can also be identified with permutations of factors in the tensor product. This is how the Schur-Weyl duality follows from Howe duality. Now let us look at how
the action of operators $t_{ij}= e_{ij}e_{ji}$ on $V_\mu^{SL_N}[0]$ translates to the action on the tensor
product when $\mu$ is a partition of $N$.

Computing in the Gelfand-Zeitlin basis it is easy to show that when $\mu$ is a partition of $N$
\[
t_{ij}=1+P_{ij}
\]
where $P_{ij}$ is the permutation of $i$-th and $j$-th factors in the tensor
product.
Note that  the identity
\[
[t_{ij}+t_{kj}, t_{ik}]=0
\]
holds in $V_\mu[0]$ for any $\mu$.

In this case the Hamiltonians of quantum spin Calogero-Moser model can be written as
\[
H=-\Delta +\sum_{i<j} \frac{l(l+P_{ij})}{\sin(x_i-x_j)^2}
\]
which acts in $L^2(\RR^n)\otimes (\CC^n)^{\otimes N}$ with $l=1$. This is the
family of Hamiltonians in physics literature as quantum spin Calogero-Moser systems,
see for example \cite{HW}.

\subsection{} When $\mu$ is a one row partition, the subspace $V_\mu[0]$ is non-zero only if $\mu=Nk$.
In this case $t_{ij}$ act by multiplication on the same scalar (independent of $i$ and $j$).
The value of this scalar is easy to compute from the action of the Casimir element
and we have in this case
\[
t_{ij}=2k(k+1)
\]
which corresponds to the usual, non-spin quantum Calogero-Moser system.
In this case $\dim( Hom(V_\lambda\otimes V_\lambda, V_{m\omega_1}))=1$.


\begin{thebibliography}{99}


\bibitem{ABB} J.Avan, O.Babelon and E.Billey. The Gervais-Neveu
-Felder Equation and the
Quantum Calogeor-Moser System.
Comm. in Math. Physics
178:281-299(1996)

\bibitem{AF}  Superintegrability of rational Ruijsenaars-Schneider systems and their action-angle duals
V. Ayadi, L. Feher, T.F. Gorbe J. Geom. Symmetry Phys. 27 (2012) 27-44.

\bibitem{C} F. Calogero, Solution of the one-dimensional N-body problem with quadratic and/or in-
versely quadratic pair potentials, J. Math. Phys. 12 (1971) 419-436.

\bibitem{M} Moser, J. Three integrable Hamiltonian systems connected with isospectral deformations. Advances in Math. 16 (1975), 197–220.

\bibitem{Et}  Etingof P., Kirillov A. (Jr), On a unified representatiuon theoretical approach
to special functions, Funk. Anal. i Prilozh., 28 (1994), 91-94; Etingof, P., Frenkel, I., Kirillov, A. , Spherical functions on affine Lie groups, Duke Math Journal, 80 (1995), 79-90; Etingof, P.I.,
Quantum integrable systems and representations of Lie alge
bras, Journal of Mathematical Physics v. 36 (1995) pp.2636-2651.

\bibitem{EV}
Etingof, Pavel, Varchenko, Alexander Traces of intertwiners for quantum groups and difference equations. I. Duke Math. J. 104 (2000), no. 3, 391–432;Etingof, P., Schiffmann, O., Varchenko, A. Traces of intertwiners for quantum groups and difference equations. Lett. Math. Phys. 62 (2002), no. 2, 143–158; Etingof, P., Varchenko, A. Orthogonality and the QKZB-heat equation for traces of Uq(g)-intertwiners. Duke Math. J. 128 (2005), no. 1, 83–117.


\bibitem{FP}  Feher, L.; Pusztai, B. G. Twisted spin Sutherland models from quantum Hamiltonian reduction. J. Phys. A 41 (2008), no. 19, 194009

\bibitem{FM} A.T. Fomenko. Symplectic geometry. Advanced Studies
in Contemporary Mathemat-
ics, 5. New York: Gordon and Breach Science Publishers, 1988

\bibitem{Winter}
J.~Frish, V.~Mandrosov,Y.A.~Smorodinsky, M.~Uhlir and P.
~Winternitz.
\newblock On higher symmetries in quantum mechanics
\newblock {\em Physics Letters} 16:354-356 (1965)

\bibitem{GH} Gibbons J., Hermsen T., A generalization of the Calogero
-Moser system., Physica, 11D(1984), 337

\bibitem{HW} K. Hikami, M. Wadati, Integrability of Calogero-Moser spin system,
Journal of the Physics Society of Japan, v. 62, n. 2 (1993) 469-472.

\bibitem{KKS} D. Kazhdan, B. Kostant and S. Sternberg. Hamiltonia
n group actions and dynamical systems of Calogero type.
Comm. Pure Appl. Mathe
31:n4, 481-507(1978)

\bibitem{KBBT} I. Krichever, O. Babelon, E. Billey and M. Talon, Spin Generalization of
Calogero-Moser system and the matrix KP equation. Translations of AMS, series 2, v. 170, Advances in Mathematical Science, Topics in Topology and Math. Phys., 1975,hep-th/9411160.

\bibitem{Ku} Kuznetsov, Vadim B., Hidden symmetry of the quantum Calogero-Moser system. Phys. Lett. A 218 (1996), no. 3-6, 212–222.

\bibitem{LX} L.C. Li, P. Xu, Spin Calogero-Moser systems associat
ed with simple Lie algebras
C.R.Acad. Sci. Paris, Serie I
, 331: n1, 55-61(2000)

\bibitem{NK} Ryota Nakai, Yusuke Kato,Particle Propagator of Spin Calogero-Sutherland Model
     J. Phys. A: Math. Theor. 47 (2014) 305205

\bibitem{N} N.N. Nekhoroshev. Action-angle variables and their g
eneralizations.
Trans. Moscow
Math. Soc.
26:180-197 (1972)

\bibitem{OP} M.A. Olshanetsky and A.M. Perelomov, Quantum integrable systems related to Lie algebras,
Phys. Rept. 94 (1983) 313-404

\bibitem{P} W. ~Pauli, On the hydrogen spectrum from the standpoint of the new quantum mechanics, Zeitschrift fur Physik, 36, 336-363 (1926).


\bibitem{R}  N. Reshetikhin, Degenerate integrability of the spin Calogero-Moser systems and the duality with the spin Ruijsenaars systems. Lett. Math. Phys. 63 (2003), no. 1, 55–71.

\bibitem{R1} N. Reshetikhin, Degenerately Integrable Systems, preprint, arXiv, 2015.

\bibitem{SInt} Superintegrability in Classical and Quantum Systems, Edited by: P. Tempesta, P. Winternitz, J. Harnad, W. Miller, Jr., G. Pogosyan, M. Rodriguez, CRM Proceedings and Lecture Notes, Volume: 37, 2004.

\bibitem{Su} B. Sutherland, Exact results for a many-body problem in one dimension. II. Phys. Rev. A, v. 5, n 3 (1972), 1372-1376.

\bibitem{W} S. Wojciechowski, 1983 Superintegrability of the Calogero-Moser system
Phys.Lett.A, 95, 279





\end{thebibliography}
\end{document}